\documentclass[12pt]{article}
\begin{document}

\setlength{\unitlength}{1mm}
%\baselineskip 0.1cm
%\large

%\setcounter{page}{1}
%\pagenumbering{arabic}
 \title{Wormhole and C-field.}
 \author{\Large $F.Rahaman^*$,$B.C.Bhui^{**}$ and P.Ghosh.}
\date{}
 \maketitle
 \begin{abstract}
                                  It is well known that a C-field, generated by a certain source
equation leads to interesting changes in the cosmological
solutions of Einstein's equations. In this article we present and
analyze a simple Lorentzian vacuum wormhole in the presence of
C-field.
  \end{abstract}

  %\bigskip
 %\medskip

 \bigskip
 \medskip
  \footnotetext{ Pacs Nos : 04.20.Gz ; 04.50.+h \\
     \mbox{} \hspace{.2in} Key words and phrases  : Wormhole, C-field\\
                              $*$Dept.of Mathematics,Jadavpur University,Kolkata-700 032,India\\
$**$ Dept. of Maths., Meghnad Saha Institute of Technology,
                                       Kolkata-700150, India\\
     E-Mail:farook\_ rahaman@yahoo.com
                              }

    \mbox{} \hspace{.2in}  The wormhole is very interesting
    subject in modern cosmology since Morris and Throne have
    verified the realistic possibilities of constructing a
    traversable wormhole space-time and traveling through it in
    the theoretical context of the general relativity [1]. Among the reasons that support this, one of them is the possibility of constructing time machines and another is related on the requirement of matter violating the weak energy condition [2].
    Wormholes may defined as handles or bridges linking different
    universes or widely separated regions of our universe.
    Topologically, wormhole space-times are the same as that of
    blackholes, but a minimal surface called throat of wormhole is
    maintained in time evolution and then a traveler can pass
    through it in both directions. To hold such a wormhole open,
    the stress energy tensor of matter violets the null energy
    conditions. As a result, the energy density of matter may be
    seen as negative by some observer. There are different ways of
    evading these violations. Most of these attempts focus on
    alternative gravity theories or existing of exotic matter [3].

        We consider the wormhole in presence of C-field. Existence of Big-bang singularity is one of the basic failures of general theory of relativity. So alternative theories are being proposed time to time. One of the important alternative theory is C - field theory introduced by Hoyle and Narlikar (HN)[4]. HN adopted a field theoretic approach introducing a mass-less and charge-less scalar field C in the Einstein - Hilbert action to account for the matter creation. A C - field generated by a certain source equation, leads to interesting changes in the cosmological solution of Einstein field equations. The modified Einstein equation due to HN through the introduction of an external
C - field are

      \begin{equation}
              R^{ab} - \frac{1}{2}g^{ab}R
             = - 8\pi G [T^{ab}- f{ C^aC^b +
             \frac{1}{2}fg^{ab}C^iC_i}]
           \end{equation}

      where C, a scalar field representing creation of matter, $x_i$ , i = 0,1,2,3 stand for the

space-time coordinates with $C_i = \frac{\partial C}{ \partial
x_i}$ and $T_{ab}$ is the matter tensor and $f
>  0$. The conservation equation in the general case is given by
  \begin{equation}
             T^{ab}_{;b}= f[ C^aC^b -
             \frac{1}{2}C^iC_i]_{;b}
           \end{equation}
 According to Sach et al [5],
the above equation could be understood as representing two
different modes of the evolution: firstly when both sides of the
equation are individually zero is non-creative mode and secondly
when are both equal and non-zero is creative mode.We study only
on the non-creative mode solution. Let us consider the static
spherically symmetric metric as
\begin{equation}
               ds^2=  - e^\nu dt^2+ e^\mu dr^2+r^2( d\theta^2+sin^2\theta
               d\phi^2)
         \label{Eq3}
          \end{equation}

The independent field equations for the metric (3) are
\begin{equation}e^{-\mu}
[\frac{1}{r^2}-\frac{\mu^\prime}{r}]-\frac{1}{r^2}= 4\pi
Gfe^{-\mu}{(C^\prime)^2}\end{equation}
\begin{equation}e^{-\mu}
[\frac{1}{r^2}+\frac{\nu^\prime}{r}]-\frac{1}{r^2}= -4\pi
Gfe^{-\mu}{(C^\prime)^2}\end{equation}
\begin{equation}e^{-\mu}
[\frac{1}{2}(\nu^\prime)^2+ \nu^{\prime\prime}
-\frac{1}{2}\mu^\prime\nu^\prime + \frac{1}{r}({\nu^\prime-
\mu^\prime})] = 8\pi Gfe^{-\mu}{(C^\prime)^2}\end{equation} Now
following the expressions (4) -(5) - (6), we have
\begin{equation}\frac{1}{2}\mu^\prime\nu^\prime
=2\frac{\nu^\prime}{r}+\frac{1}{2}(\nu^\prime)^2+\nu^{\prime\prime}\end{equation}
One may attempt to eq.(7) for two different situations:
(i)$(\nu^\prime)= 0$, which means $ \nu=constant$.
(ii)$(\nu^\prime)\neq 0$, which in turn leads to the differential
equation $ e^\nu(\nu^\prime)^2=\frac{b^2e^\mu}{r^4}$, where b is
an arbitrary integrating constant. In the first case, one may
choose $ \nu =0$ , without any loss of generality. In the second
case i.e. for $(\nu^\prime)\neq 0$, it is extremely hard task to
obtain the solutions to the field equations. Hence we abandon
this case in our present work and proceed with the other case
i.e.$(\nu^\prime)= 0$ for further calculations. With
$(\nu^\prime)= 0$, one gets, after some simple and straight
forward calculations the following two equations:
\begin{equation}e^{-\mu}
[\frac{2}{r^2}-\frac{\mu^\prime}{r}]= \frac{2}{r^2}\end{equation}
\begin{equation}\frac{(e^{-\mu}) ^\prime }{r} =8\pi Gfe^{-\mu}{(C^\prime)^2}\end{equation}
Solving the above two equations, we get
\begin{equation}e^{-\mu}= 1-\frac{D}{r^2}\end{equation}
\begin{equation}C=\frac{1}{\sqrt(4\pi
Gf)}\sec^{-1}\frac{r}{\sqrt(D)}+ C_0\end{equation} where D and
$C_0$ are integration constants. Thus our solution reads
\begin{equation}
               ds^2=  - dt^2+ [ 1-\frac{D}{r^2}] ^{-1}dr^2+r^2( d\theta^2+sin^2\theta
               d\phi^2)
          \end{equation}
In order to investigate whether a given solution represents a
wormhole geometry, we study the geodesic. The equation of
geodesic for the metric given in (12)
\begin{equation}r^{\**2}\equiv[\frac{dr}{ds}]^2 = [ 1 - \frac{D}{r^2}][ E^2 - \frac{J^2}{r^2}- L]\end{equation}
\begin{equation}\phi^{\**}\equiv\frac{d\phi}{ds} =\frac{J}{r^2}   \end{equation}
\begin{equation}t^{\**}\equiv\frac{dt}{ds} =E   \end{equation}
where the notion is as usual considered to be taking place in the
$\theta = \frac{\pi}{2}$  planes and the constants E and J having
the respective interpretations of energy per unit mass and angular
momentum about an axis perpendicular to the invariant plane
$\theta = \frac{\pi}{2}$. Here s is an affine parameter and L is
the Lagrangian having values 0 and 1 respectively for null and
time like particles. \linebreak Now the equation for radial
geodesics $( J = 0 )$:
\begin{equation}[\frac{dr}{ds}]^2 = [ 1 - \frac{D}{r^2}][ 1 - \frac{L}{E^2}]\end{equation}
with solution
\begin{equation}t=\pm\sqrt (\frac{ [ r^2 -D]}{[ 1 -
\frac{L}{E^2}]} ) + constant\end{equation} and the affine
parameter $ s  \alpha  t $. The eq.(21) represents a hyperbola and
shows that to an external observer a radially in-falling time
like or null particle approaches the radius $r
=\sqrt(D)$asymptotically but can never reach it. Here we also see
the s - r relationship represents a hyperbola. Now, for time like
geodesics $( L = 1 )$, s is the proper time and hence an observer
falling with time like particle also skirts the physical
singularity at $r = 0$ by asymptotically grazing the critical
radius at $r =\sqrt(D)$.This feature is characteristic of a
wormhole space-time. Now we rewrite the metric into Morris-Thorne
cannonical form:
\begin{equation}
               ds^2=  - dt^2+ \frac{dr^2}{[ 1-\frac{b(r)}{r}]}+r^2( d\theta^2+sin^2\theta
               d\phi^2)
          \end{equation}
Here $ b(r) =\frac{D}{r} $ is called shape function. This
function should satisfy some constraint, enumerated in [1,2]. The
throat of the wormhole occurs at $ r=r_0=\sqrt(D)>0$. We notice
that since now $r\geq r_0 > 0$, there is no horizon. Here one can
see that $ \frac{b(r)}{r}\leq1 $ and $ \frac{b(r)}{r}\ $ tends to
zero as r tends to infinity. Moreover, it results that $
\frac{b(r)}{r}=1$ when r reaches its minimum value $r_0$. Thus
the function b(r) meets all the requirements needed to describe
what can be called a wormhole. The line element for an equatorial
slice through the wormhole at a fixed instant of time is
\begin{equation}
               ds^2=   [ 1+(\frac{dz}{dr}) ^2] dr^2+r^2( d\theta^2+sin^2\theta
               d\phi^2)
          \end{equation}

where the embedding function z(r) is a solution of
\begin{equation}
               \frac{dz}{dr}= \pm\sqrt \frac{1}{[\frac{r}{b(r)}
               -1]}\end{equation}
At the value $ r = r_0$ ( the wormhole throat) eq.(20) is
divergent, which means that the embedded surface is vertical
there. For a coordinate independent description of wormhole
physics, one may use proper length $'l'$ instead of $'r'$ such
that
\begin{equation}
               l = +\int_{r_0^+}^{r}\frac{dr}{\sqrt(1-\frac{b(r)}{r}
               )}\end{equation}
Thus $ l=\pm\sqrt ( r^2 -D)$. Due to the simple expression for
$l(r)$ it is to rewrite the metric tensor interms of this proper
radial distance
\begin{equation}
               ds^2=  - dt^2+ dl^2+r^2(l) [ d\theta^2+sin^2\theta
               d\phi^2]
          \end{equation}
where $l^2=( r^2 -D)$. Thus in this well behaved coordinate
system, as $l$ increases from $- \infty  to  0$,r decreases
monotonically to a minimum value at the throat; and as $l$
increases onwards to $+\infty$, r increases monotonically. It can
be verified that all the conditions  of a two way wormhole
including the flaring-out conditions are satisfied. The
pecularity of this solution is that
\begin{equation}\frac{db}{dr} =-\frac{D}{r^2}<0 \end{equation}
and hence $G_{00} < 0$ for all finite nonzero values of $'r'$.
This implies that the entire wormhole, and not only the throat, is
made up of exotic material.\linebreak Summing up, we showed that
C-field admits analytical wormhole solutions. It should be noted
that there exists some regions in which C-field may play the role
of exotic matter. This implies that it might be possible to build
a wormhole like space-time with the presence of ordinary matter at
the throat.

        { \bf Acknowledgements }

        We are grateful to Dr.A.A.Sen for helpful discussions. F.R is also thankful
        to IUCAA for providing research facility.\\

%\begin{figure}[p]
%\includegraphics*[450,350]{fig1.bmp}
%\caption{Variation of deflection of the circular plate}
%\end{figure}

\end{document}